\documentclass[twocolumn,showpacs,preprintnumbers,amsmath,amssymb]{revtex4-1}
\usepackage{graphicx}
\usepackage{dcolumn}
\usepackage{bm}

\newcommand{\be}{\begin{eqnarray}}
\newcommand{\ee}{\end{eqnarray}}

\begin{document}


\title{Transverse instability and disintegration of domain wall of relative phase in coherently coupled two-component Bose-Einstein condensates}

\author{Kousuke Ihara}
\author{Kenichi Kasamatsu}
\affiliation{Department of Physics, Kindai University, Higashi-Osaka, Osaka 577-8502, Japan}

\date{\today}

\begin{abstract}
We study transverse instability and disintegration dynamics of a domain wall of a relative phase in two-component Bose-Einstein condensates 
with a coherent Rabi coupling. We obtain analytically the stability phase diagram of the stationary solution of the domain wall for the one-dimensional coupled Gross-Pitaevskii equations 
in the plane of the Rabi frequency and the intercomponent coupling constant. 
Outside the stable region, the domain wall is dynamically unstable for the 
transverse modulation along the direction perpendicular to the phase kink. The nonlinear evolution associated with the instability is demonstrated through numerical simulations 
for both the domain wall without edges and that with edges formed by the quantized vortices. 
\end{abstract}

\maketitle

\section{Introduction} \label{intro}
Solitons in multidimensional systems are generally unstable, known as the transverse instability \cite{kivshar1998dark,carr2008multidimensional}, 
where the solitonic structure is dynamically unstable against the symmetry-breaking modulation along the unbounded spatial dimension. 
Bose-Einstein condensates (BECs) are well-suited system to study such an unstable properties of solitons. This is because: (i) 
The properties of BECs are described by the Gross-Pitaevskii (GP) (nonlinear Sch\"{o}redinger) equation, which allows various solitary wave solutions. (ii) The system consists of dilute gases and is almost far from dissipation so that nonequilibrium processes 
of unstable dynamics can be directly observed from the well-defined initial states. (iii) The topological solitons can be created by well-developed phase-engineering techniques. 
Given all three of these advantages, dynamical instabilities of solitons have indeed been observed in dilute BECs.
For example, the dark solitons in two-dimensional (2D) BECs are broken up 
to an array of vortex pairs due to the transverse instability, called ``snake instability" \cite{PhysRevA.60.R2665,PhysRevA.65.043612}. The decay of a planer dark 
soliton into a vortex ring was observed experimentally in a 3D BEC \cite{PhysRevA.62.053606,PhysRevLett.86.2926}. 

For two-component BECs described by the two-component order parameters, 
the structures of solitons are richer than those in single-component BECs, e.g., dark solitons \cite{PhysRevLett.86.2918}, dark-bright solitons \cite{PhysRevLett.87.010401,becker2008oscillations}, 
domain walls \cite{PhysRevLett.87.140401}, and magnetic solitons \cite{PhysRevLett.116.160402}. 
For a typical two-component mixture of BECs, the U(1) phases in each component are independent variables, 
because the two components are coupled only through the density-density coupling. 
However, when a Rabi coupling is applied between the internal states of two-component BECs, the symmetry associated with 
one of the two U(1) phases is broken, and the relative phase between the two components makes sense. 
Then, the new type of soliton can exist as ``a domain wall of the relative phase" which is obtained as 
a kink solution of the sine-Gordon equation for the two-component BECs \cite{son2002domain}. 
This domain wall can exist as a bound string between two vortices in 
two-component BECs, which results in a molecule of the vortices \cite{kasamatsu2004vortex}. 

Recently, real time dynamics of the aforementioned vortex molecule in the Rabi-coupled condensates have been 
studied by some authors \cite{tylutki2016confinement,calderaro2017vortex,eto2018confinement}. The authors in Refs.\cite{tylutki2016confinement,eto2018confinement} have especially stressed that this system can be 
used as the analogous simulations of the ``confinement" in quark--anti-quark phenomena. 
Tylutki \textit{et al}. studied the precession dynamics of the vortex molecule as a function 
of the Rabi coupling and the molecular distance \cite{tylutki2016confinement}. 
In a certain regime, the domain wall was found to disintegrate into some parts, which is analogous to 
the string breaking phenomena in quantum chromodynamics. 
This observation was further confirmed by the simulations by Eto and Nitta \cite{eto2018confinement}. 

Here, we reveal the physical origin of this disintegration as the instability associated with the transverse displacement of the domain wall, i.e, the snake instability. 
We obtain an exact solution of the domain wall for the 1D coupled GP equations for two-component BECs, making a phase diagram of the stationary 
solutions in the plane of the intercomponent coupling strength and the Rabi frequency.
The energetic stability of the solution reproduces the previous work by 
Usui and Takeuchi \cite{usui2015rabi}. In the unstable regime, the dynamical instability 
takes place for the transverse modulation along the direction perpendicular to the sine-Gordon kink, which is 
analyzed by the Bogoliubov-de Genne (BdG) equation. The nonlinear dynamics associated with 
this instability are shown by direct numerical simulations 
of the 2D GP equation for the domain wall without edges and with edges formed by quantized vortices. 
The disintegration observed in previous works \cite{tylutki2016confinement,eto2018confinement} can be explained as the manifestation of 
the snake instability.

The paper is organized as follows. After introducing the formulation of the problem in Sec.~\ref{formulation}, 
we first study the stability of the domain wall of the relative phase based on the exact solution of the GP equation in Sec.~\ref{exactsolsec}. 
Next, the transverse instability of the domain wall is discussed through the BdG analysis in Sec.~\ref{BdGAnaly}. 
In Sec.~\ref{numericalsim} we demonstrate the dynamics due to the transverse instability using the direct numerical simulations of the 
2D GP equations. Section~\ref{condle} is devoted to the conclusion. 

\section{formulation}\label{formulation}
We consider two-component BECs of ultracold atoms having the same mass $m$ and residing in two different hyperfine states. 
The BECs are described by the condensate wave functions $\Psi_{j}$ ($j=1,2$). 
The equilibrium state of the system is obtained by minimizing the Gross-Pitaevskii (GP) energy functional
\begin{eqnarray}
E[\Psi_{1},\Psi_{2}] = \int d \bm{r} \biggl[ \sum_{j=1,2} \biggl( \Psi_{j}^{\ast}  h \Psi_{j} 
+ \frac{g_{j}}{2} |\Psi_{j}|^{4}  \biggr)  \nonumber \\
+ g_{12} |\Psi_{1}|^{2} |\Psi_{2}|^{2} 
-\frac{\hbar \Omega_{\rm R}}{2} (\Psi_{1}^{\ast} \Psi_{2} + \Psi_{1} \Psi_{2}^{\ast} ) \biggr].
\label{coupleGPfunc} 
\end{eqnarray}
Here, $h = -\hbar^2 \nabla^{2}/(2m) + V_\text{ext} $ is the single-particle Hamiltonian with the 
trapping potential $V_\text{ext}$. 
The coupling constants $g_j$ ($j=1,2$) and $g_{12}$ represent the strength of intra- and the intercomponent interactions, respectively, 
described as $g_j = 4 \pi \hbar^2 a_j / m$ and $g_{12} = 4 \pi \hbar^2 a_{12} / m$ with the $s$-wave scattering lengths $a_i$ and $a_{12}$
between the corresponding atoms. We assume that the intracomponent coupling constants satisfy $g_1 = g_2 = g > 0$ for simplicity. 
The last term in Eq.~(\ref{coupleGPfunc}) describes a coherent Rabi coupling induced by an external electromagnetic field, which allows atoms to transfer 
their internal states coherently \cite{hall1998measurements,matthews1999watching}; $\Omega_{\rm R}$ stands for the Rabi frequency. Then, the total number $N=N_1+N_2 = \int d\bm{r} (|\Psi_1|^2+|\Psi_2|^2)$ is a constant of motion. 
In the following analysis except the numerical simulations in Sec.~\ref{cylinsim}, we assume the homogeneous system by setting $V_\text{ext} = 0$. 
For the coupling constants, we confine ourselves to the range $-1<g_{12}/g<1$ corresponding to a miscible regime without the Rabi coupling, 
otherwise the components phase separate for $g_{12} > g$ \cite{timmermans1998phase,ao1998binary} or 
undergo mean-field collapse for $g_{12} < -g$ \cite{adhikari2001stability}. The properties of a domain wall of the condensate densities for $g_{12} > g$ in the presence of the Rabi-coupling 
have been studied in detail in Ref.~\cite{dror2011domain}.

The time-dependent GP equations are given by the variational procedure $i \hbar \partial \Psi_j / \partial t = \delta E /\delta \Psi_j^{\ast}$ as 
\begin{align}
i \hbar \frac{\partial \Psi_1}{\partial t} &=  - \frac{\hbar^2 \nabla^2}{2m} \Psi_1 
 + g |\Psi_1|^2 \Psi_1 + g_{12} |\Psi_2|^2 \Psi_1 - \frac{\hbar \Omega_\text{R}}{2} \Psi_2,  \label{GPtwo1}\\
i \hbar \frac{\partial \Psi_2}{\partial t} &=  - \frac{\hbar^2 \nabla^2}{2m} \Psi_2 
+ g |\Psi_2|^2 \Psi_2 + g_{12} |\Psi_1|^2 \Psi_2 - \frac{\hbar \Omega_\text{R}}{2} \Psi_1. \label{GPtwo2}
\end{align}
First, we consider the ground state in the homogeneous system by ignoring the time and spatial derivative terms in Eqs.~\eqref{GPtwo1} and \eqref{GPtwo2}. 
Because of the conservation of the total particle number, the chemical potential as a Lagrange multiplier for 
both components should be common as 
$\mu_1 = \mu_2 = \mu$, the stationary wave function being written as $\Psi_j = \sqrt{n_j} e^{i \theta_j - i \mu t /\hbar}$. Then, the relative phase $\theta = \theta_1 - \theta_2$ should be vanished, because the energy of the Rabi-coupling 
term $- \hbar \Omega_\text{R} \sqrt{n_1 n_2} \cos \theta$ is minimized at $\theta = 0$. 
As a result, the densities in the miscible regime satisfy $n_1 = n_2 \equiv n_0$ with 
\begin{equation}
n_0 =\frac{\mu + \hbar \Omega_R/2}{g+g_{12}}.
\end{equation}
The miscible regime takes place when the intercompnent coupling constant satisfies $g_{12} < g + \hbar \Omega_R/(n_1+n_2)$ \cite{abad2013study}, 
otherwise the equilibrium state involves spontaneous density imbalance. 

We scale the wave function as $\Psi_j = \sqrt{n_0} \psi_j$, and introduce the length, time, and energy scale as 
$\xi = \hbar/\sqrt{2m g n_0}$, $\hbar/(g n_0)$, and $g n_0$, respectively. 
Then, we get the dimensionless GP equation, 
\begin{align}
i  \frac{\partial \psi_1}{\partial t} =  - \nabla^2 \psi_1 - \tilde{\mu} \psi_1 + |\psi_1|^2 \psi_1 + \gamma |\psi_2|^2 \psi_1 - \omega_\text{R} \psi_2, \label{GPlesstwo1} \\
i \frac{\partial \psi_2}{\partial t} =  - \nabla^2 \psi_2 - \tilde{\mu} \psi_2 + |\psi_2|^2 \psi_2 + \gamma |\psi_1|^2 \psi_2 -  \omega_\text{R} \psi_1.  \label{GPlesstwo2}
\end{align}
Here, all the variables are dimensionless and the coefficients are given as
\begin{align}
\omega_\text{R} = \frac{\hbar \Omega_\text{R}}{2 g n_0}, \quad \gamma = \frac{g_{12}}{g}, \quad \tilde{\mu} = 1+\gamma-\omega_\text{R}.
\end{align}

\section{Transverse instability of domain wall of relative phase}\label{linearanalysis}
In the seminal paper \cite{son2002domain}, Son and Stephanov showed that the GP equation for the Rabi-coupled two-component BECs 
can be reduced to the sine-Gordon equation when the gradient of the density is neglected. 
Considering the 1D system along the $x$-axis and substituting the expression 
$\psi_j (x) = e^{i \theta_j (x)}$ into Eqs.~\eqref{GPlesstwo1} and \eqref{GPlesstwo2}, we get the sine-Gordon equation 
\begin{equation}
\frac{\partial^2 \theta}{\partial x^2} = 2 \omega_\text{R} \sin \theta
\end{equation}
with the relative phase $\theta \equiv \theta_1 - \theta_2$ and the constant $C$ of integral. 
The stationary solution for the boundary conditions $\theta = \pi \pm \pi$ and $\partial_x \theta = 0$ for $x \to \pm \infty$ is given by 
\begin{equation}
\theta (x) = 4 \arctan e^{\sqrt{2 \omega_\text{R}} x}.  \label{sGsolitonform}
\end{equation}
This solution is known as the sine-Gordon soliton \cite{son2002domain}. 
The stability criterion was obtained as $\omega_\text{R}^{c} = (1-\gamma)/3$ by neglecting 
the density gradient, which is valid for $\gamma \sim 1$.
Son and Stephanov proposed that, when vortices are present in two-component BECs, the Rabi coupling can bind a pair of vortices in 
the different components via the sine-Gordon domain wall. 
After that the binding of vortices and composite structures of the resulting vortex molecules 
have been studied in various situations 
\cite{kasamatsu2005spin,kasamatsu2005vortices,eto2012vortex,cipriani2013crossover,dantas2015bound,aftalion2016rabi,uranga2018infinite,tylutki2016confinement,eto2018confinement,calderaro2017vortex,shinn2018mesoscopics}. 

In this section, we consider the stability of the domain wall of the relative phase based on the GP equations \eqref{GPlesstwo1} and \eqref{GPlesstwo2}, 
where we take account of the contribution of the density gradient. We first consider the energetic stability of the 
domain wall. Although this problem has been considered by several papers \cite{son2002domain,usui2015rabi,qu2017magnetic}, 
we employ the exact solution of the domain wall and extend the phase diagram 
to the region of the negative intercomponent coupling $\gamma < 0$. 
Next, we consider the transverse instability by extending the 1D domain wall to the additional spatial dimension. 
The stability can be studied by the BdG analysis, where the signal of the 
instability is shown by the appearance of the imaginary excitation frequency of the Bogoliubov modes. 

\subsection{Exact solution and energetic instability}\label{exactsolsec}
\begin{figure}[ht]
\centering
\includegraphics[width=0.9\linewidth]{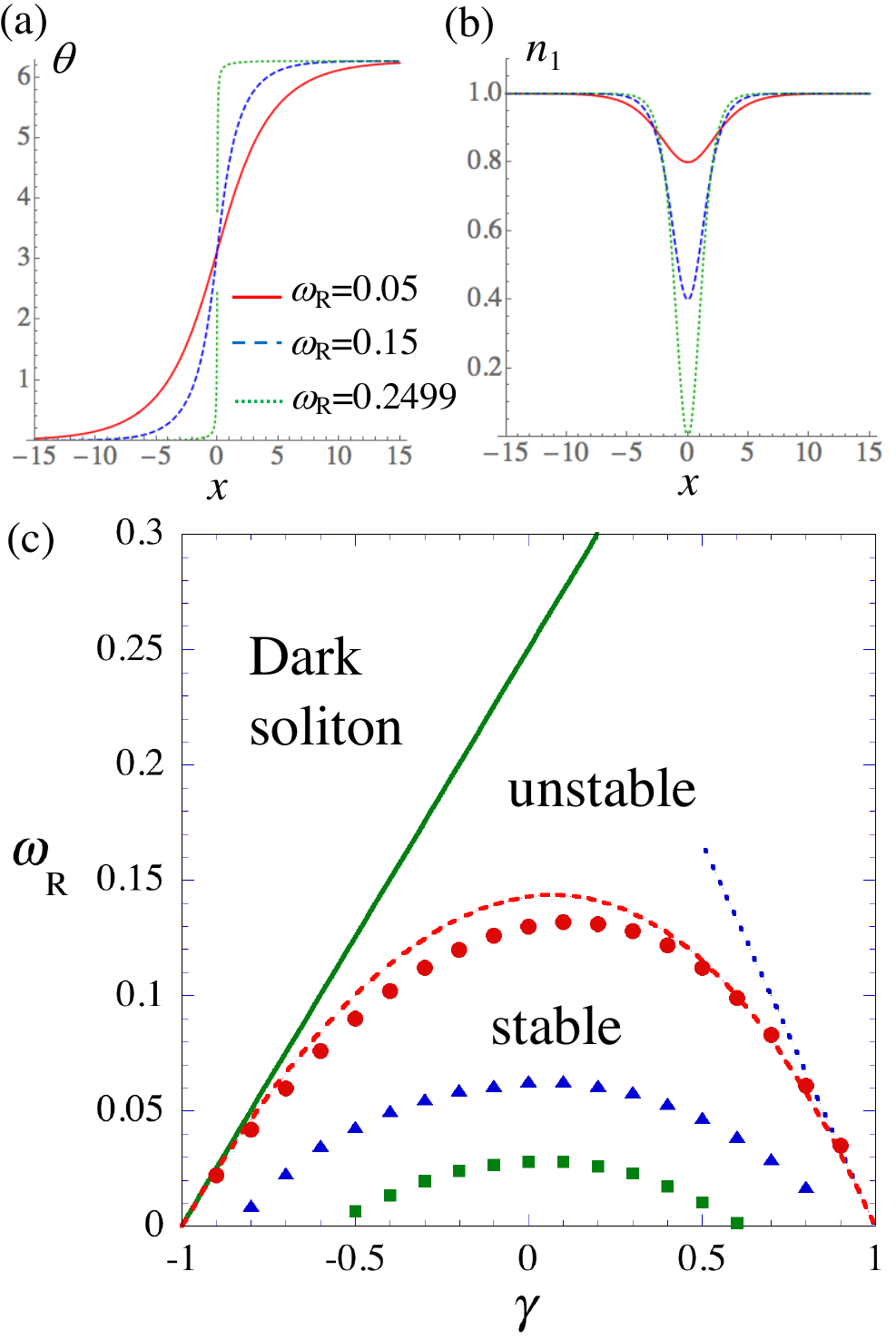}
\caption{The upper panels show the profile of the relative phase $\theta = \theta_1 -\theta_2$ (a) and the density $n_1 = n_2$ (b) of the exact solutions for $\gamma = 0$ and several values of $\omega_\text{R}$. The lower panel (c) represents the phase diagram of the sine-Gordon domain wall in the $\gamma$-$\omega_\text{R}$ plane. 
The red filled circles represent the stability boundary obtained by the imaginary time propagation of the GP equations. 
The green solid line, the red dashed curve, and the blue dotted line represent $\omega_\text{R} = \omega_\text{R}^{u}$, $\omega_\text{R} = \omega_\text{R}^{c}$ by Eq.~\eqref{usuit}, and $\omega_\text{R} = \omega_\text{R}^{c}$ in Ref.~\cite{son2002domain}, respectively. 
The blue triangles and green squares are stability boundary for the moving domain wall with $V=0.2$ and $V=0.4$, respectively, discussed in Sec.~\ref{cylinsim}.}
\label{sgphased}
\end{figure}

We consider the energetic stability of the domain wall of the relative phase by solving the 1D version of the GP 
equations \eqref{GPlesstwo1} and \eqref{GPlesstwo2}. 
We seek the stationary solutions which satisfy $|\psi_j| = 1$ at infinity and the phases change continuously as $0 \to + (-) \pi$ for $\psi_1$ ($\psi_2$) from $x=-\infty$ to $x=+\infty$. 
For the sine-Gordon soliton and our symmetric parameters, the solution should satisfy the relation $\psi_1 = \psi_2^{\ast}$. 
This restriction reduces the equations to 
\begin{equation}
-\frac{\partial^2 \psi_1}{\partial x^2} - \tilde{\mu} \psi_1 + (1+\gamma) |\psi_1|^2 \psi_1 - \omega_\text{R} \psi_1^{\ast} = 0, \label{redgPeq}
\end{equation}
which allows us to get an expression of the exact solution of Eqs.~\eqref{GPlesstwo1} and \eqref{GPlesstwo2} as 
\begin{align}
\psi_1 &= \psi_2^{\ast} = -\tanh(\sqrt{2 \omega_\text{R}} x) + i A \: \text{sech} (\sqrt{2 \omega_\text{R}} x),  \label{exactsolu} \\
 A & = \sqrt{\frac{1+\gamma - 4 \omega_\text{R}}{1+\gamma}} \label{exactcoef}.
\end{align}
The typical profile of the solution is shown in Fig.~\ref{sgphased}(a) and (b). 
The relative phase changes $2 \pi$ around the origin over the length scale $\sim (2 \omega_\text{R})^{-1/2}$. 
The density profile is written as $n_1 = n_2 = \tanh^2 (\sqrt{2 \omega_\text{R}} x) + A^2 \: \text{sech}^2 (\sqrt{2 \omega_\text{R}} x)$, being uniform only for $\omega_\text{R} = 0$. With increasing $\omega_\text{R}$, the spatial profile of $\theta$ approaches to a step function and the density depression becomes deeper.
The solution of Eq.~\eqref{exactsolu} is effective below the upper bound of the Rabi frequency $\omega_\text{R}^u = (1+\gamma)/4$, at which 
the solution coincides with the form of the dark soliton with exactly zero density at the center. Physically, this boundary is 
interpreted as that the length scale $(2 \omega_\text{R})^{-1/2}$ is equal to the healing length given by the effective 
coupling constant $1+\gamma$ [see Eq.~\eqref{redgPeq}]. 

Figure~\ref{sgphased}(c) shows the phase diagram representing the energetic stability of the single domain wall 
in the $\gamma$-$\Omega_\text{R}$ plane. Below the line $\omega_\text{R} = \omega_\text{R}^{u}$, 
we have the solution of the domain wall of Eq.~\eqref{exactsolu}. Here, we check numerically the energetic stability of the solution 
Eq.~\eqref{exactsolu} through the imaginary time propagation of the GP equations \eqref{GPlesstwo1} and \eqref{GPlesstwo2}, which is plotted 
in Fig.~\ref{sgphased}(c). 
When the solution is unstable, the density difference grows and unwinds 
the 2$\pi$ difference of the relative phase to zero, leading to the uniform solution \cite{usui2015rabi}. 
The stable range of $\omega_\text{R}$ becomes narrower with increasing $|\gamma|$ and vanishes 
at $\gamma = \pm 1$. 

This result agrees with the previous works \cite{son2002domain,usui2015rabi}. 
Near  $\gamma \sim 1$, the plots are fitted $\omega^{c}_\text{R}=(1-\gamma)/3$ obtained in \cite{son2002domain}. 
Usui and Takeuchi extended the analysis 
by considering the density depression through the numerical and variational analyses \cite{usui2015rabi}. 
They proposed that the instability is associated with the Landau instability of the local counterflow across the domain wall \cite{son2002domain} 
and the critical velocity was estimated by using the local density at the depression. 
Since the instability can occur when the wavelength of the unstable excitation is smaller than 
the length scale of the domain wall, they obtained the expression for the stability as 
\begin{equation}
\omega_\text{R}^{c} = \frac{1}{3} n_\text{min} (\omega_\text{R}^{c}) (1-\gamma). \label{usuit}
\end{equation}
Here, $n_\text{min}(\omega_\text{R}^{c})$ represents the density at the density minimum for $\omega_\text{R} = \omega_\text{R}^{c}$, being written as $n_\text{min} = (1+\gamma-4 \omega_\text{R}^{c})/(1+\gamma)$ 
by using Eq.~\eqref{exactcoef}. This stability criterion is also shown in the dashed curve in Fig.~\ref{sgphased}, 
which is in good agreement with the numerical result. 

\subsection{Transverse instability of the domain wall}\label{BdGAnaly}
Next, we further study the stability of the sine-Gordon domain wall by the BdG analysis. 
Especially, we include the fluctuation along the direction ($y$-axis) perpendicular to the sine-Gordon soliton in the $x$-axis to study the transverse instability. 

In the standard BdG analysis, the wave function is expanded around the stationary solution $\psi_j^0$ as 
\begin{equation}
\psi_j = \psi_j^0 + \left[ u_j (x) e^{ i k y-i \omega t} - v_j^{\ast}(x) e^{- i k y + i \omega t} \right].
\end{equation}
Here, the fluctuation along the $y$-direction is included by the plane wave $\propto e^{iky}$ \cite{takeuchi2010quantum,suzuki2010crossover}. 
Substituting this expression into Eqs.~\eqref{GPlesstwo1} and \eqref{GPlesstwo2}, we get the eigenvalue equation. 
The eigenfrequency $\omega$ is calculated by solving the BdG equation
\begin{widetext}
\begin{align}
\hat{\cal H}{\bf u} 
= \hbar \omega {\bf u}, \quad\quad
\hat{\cal H}
=
\left( 
\begin{array}{cccc}
\hat{h}_1 & - \left( \psi_1^0 \right)^2 &\gamma \psi_1^0 \psi_2^{0\ast} - \omega_\text{R} & -\gamma \psi_1^0 \psi_2^0 \\
\left( \psi_1^{0 \ast} \right)^2 & -\hat{h}_1 & \gamma \psi_1^{0 \ast} \psi_2^{0 \ast} & -\left( \gamma \psi_1^{0 \ast} \psi_2^0 - \omega_\text{R} \right) \\
\gamma \psi_1^{0\ast} \psi_2^0 - \omega_\text{R} & -\gamma \psi_1^0 \psi_2^0 & \hat{h}_2 & - \left( \psi_2^0 \right)^2 \\
\gamma \psi_1^{0\ast} \psi_2^{0\ast} & -\left( \gamma \psi_1^0 \psi_2^{0 \ast} - \omega_\text{R} \right) & \left( \psi_2^{0 \ast} \right)^2 & -\hat{h}_2 \\
\end{array} 
\right),
\label{eq:reducedBdG}
\end{align}
\end{widetext}
where ${\bf u}=(u_1, v_1, u_2, v_2)^\text{T}$ and $\hat{h}_j= - \partial_x^2 + k^2 -\tilde{\mu} + 2  |\psi_j^0|^2 + \gamma |\psi_{\bar{j}}^0|^2$, 
where $\bar{j}=1 (2)$ for $j=2 (1)$. Employing the domain wall solution for $\psi_j^{0}$, we numerically solve Eq.~\eqref{eq:reducedBdG} with 
the finite system size $-60 \leq x \leq 60$ and the 600 grid points. To this end, starting from Eq.~\eqref{exactsolu} which is a solution in an infinite system, we make imaginary time propagation of 
the GP equation to get the proper solution $\psi_j^{0}$ for the finite-size system.

The eigenvalues of the BdG equation can be used to clarify the stability properties of the stationary solutions. 
Let us first consider the situation $k=0$, where the only 1D perturbation is present. 
In the stable region in Fig.~\ref{sgphased}, there are positive eigenfrequencies and one zero-energy mode 
when we take only eigenmodes with positive norm $\sum_j \int dx (|u_j|^2-|v_j|^2) > 0$. 
When the solution enters the unstable region in Fig.~\ref{sgphased}, 
there appears the negative eigenvalue, a signature of the energetic Landau instability. 
The magnitude of the negative eigenvalue is very small in the most of unstable region, which implies that 
the Landau instability is very weak. The imaginary time propagation confirms this feature, because 
the decay of the initial domain wall solution to the uniform one needs very long time. 
Thus, we expect that this instability is not significant in cold gas experiments at ultralow temperatures. 

\begin{figure}[ht]
\centering
\includegraphics[width=1.0\linewidth]{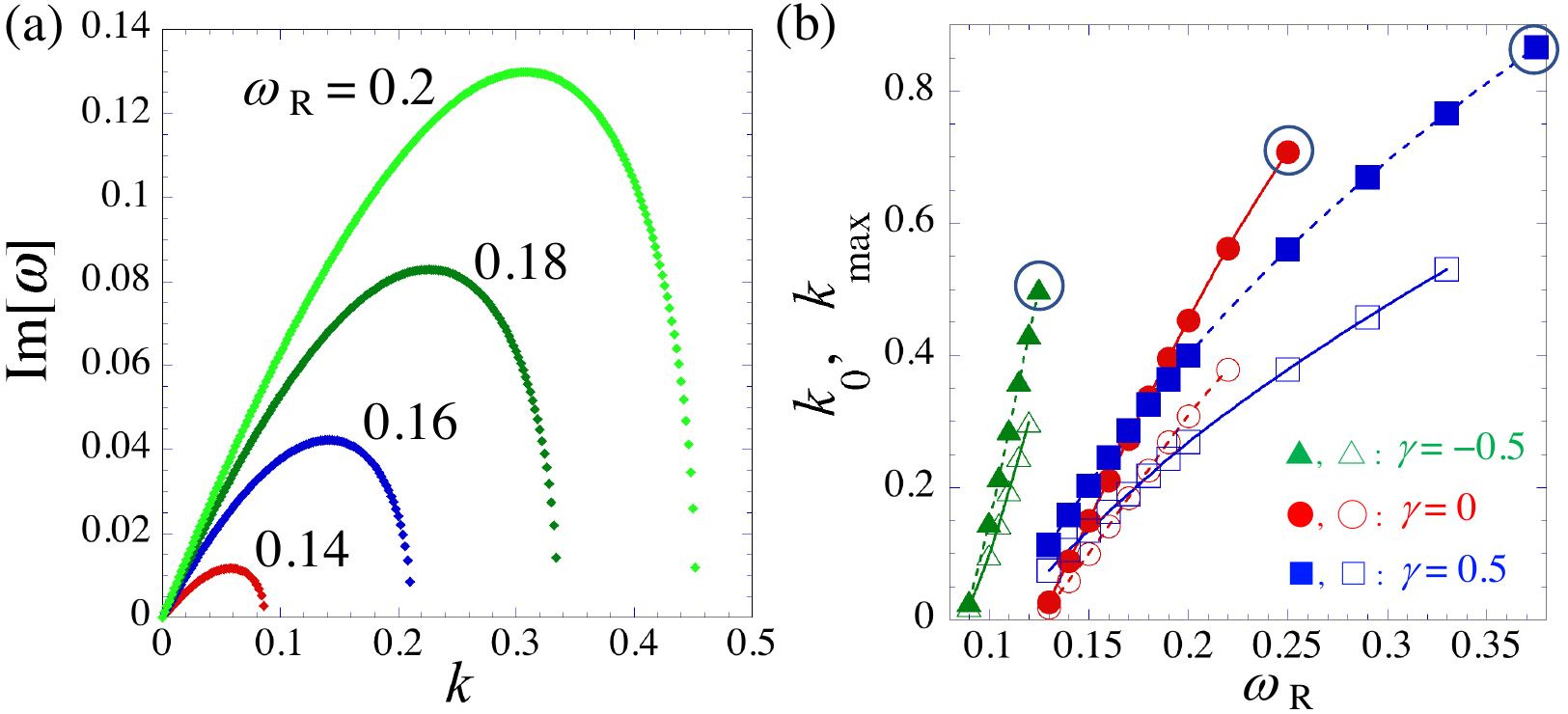}
\caption{The transverse instability of the domain wall solution Eq.~\eqref{exactsolu}. 
In (a), the imaginary part of the excitation frequency $\omega$ as a function of $k$ for $\gamma=0$ is plotted for several values of $\omega_\text{R}$. 
For a given $\omega_\text{R}$ the imaginary component appears in the region $0<k<k_\text{max}$ and 
$\text{Im}[\omega]$ takes the maximum at $k = k_0$; 
we define $k_0$ and $k_\text{max}$ as what gives the maximum of $\text{Im}[\omega]$ and the upper bound of the unstable region, respectively. 
In (b), we plot $k_0$ (open symbols) and $k_\text{max}$ (filled symbols) as a function of $\omega_\text{R}$ for $\gamma=0, \pm0.5$. The curves are interpolation as a guide. 
The symbols enclosed by circles correspond to the values of $k_\text{max}$ for a dark soliton of a single-component condensate.
}
\label{bdg}
\end{figure}
The more significant instability is the dynamical one. When we include the eigenmodes with finite $k$, 
there appear imaginary components in the excitation frequencies. 
Figure~\ref{bdg}(a) shows the imaginary part of the excitation frequency as a function of $k$ for $\gamma=0$ and several values of $\omega_\text{R}$. 
The imaginary part appears for the finite range $0 < k < k_\text{max}$ and takes a maximum value at a certain wave number $k=k_0$. 
The unstable range of $k$ is extended and the maximum of $\text{Im}[\omega]$ is enhanced with increasing $\omega_\text{R}$. 
Thus, the instability becomes stronger with increasing $\omega_\text{R}$. In Fig.~\ref{bdg}(b), we also plot $k_0$ and $k_\text{max}$ as a function of $\omega_\text{R}$ 
for $\gamma=0, \: \pm 0.5$. 
Extrapolation of the curves to $k_0 = k_\text{max}=0$ gives the critical values $\omega_\text{R}^c$ of the dynamical stability for each $\gamma$, 
which agrees with the numerically obtained $\omega_\text{R}^c$ in Fig.~\ref{sgphased}. 
We see that the parameter region exhibiting the transverse instability coincides with the unstable region of Fig.~\ref{sgphased}. 
For $\omega_\text{R} > \omega_\text{R}^u$ the solution reduces to the dark soliton, where 
the transverse instability is expected according to the previous literature \cite{PhysRevA.60.R2665,PhysRevA.65.043612}. 
As seen in Fig.~\ref{bdg}(b), the plots are connected to the values of $k_\text{max}$ for the transverse instability of the dark soliton at $\omega_\text{R} = \omega_\text{R}^u = (1+\gamma)/4$; 
we have $k_\text{max} = \sqrt{(1+\gamma)/2}$ for a dark soliton (see Eq.~(7.3) in Ref.~\cite{kivshar1998dark}) following our notation of Eq.\eqref{redgPeq}.

\section{Disintegration dynamics of a domain wall of relative phase}\label{numericalsim}
In this section, we perform numerical simulations of the GP equations to demonstrate the disintegration 
of the domain wall of the relative phase through the transverse instability. 
We consider two cases: (i) an extended domain wall without the edges in a uniform system, and 
(ii) a domain wall with a finite length in a cylindrical trap. In the latter case, the edges of 
the wall correspond to the vortices with the same circulation in each component. 
The disintegration of the domain wall has been found in the numerical simulations in Refs.~\cite{tylutki2016confinement,eto2018confinement}, 
when the separation of the vortices and the Rabi frequency are large. 
The transverse instability of a moving domain wall without edges in trapped 
BECs has been reported by C. Qu \textit{et al.} \cite{qu2017magnetic}. 

\begin{figure}[ht]
\centering
\includegraphics[width=1.0\linewidth]{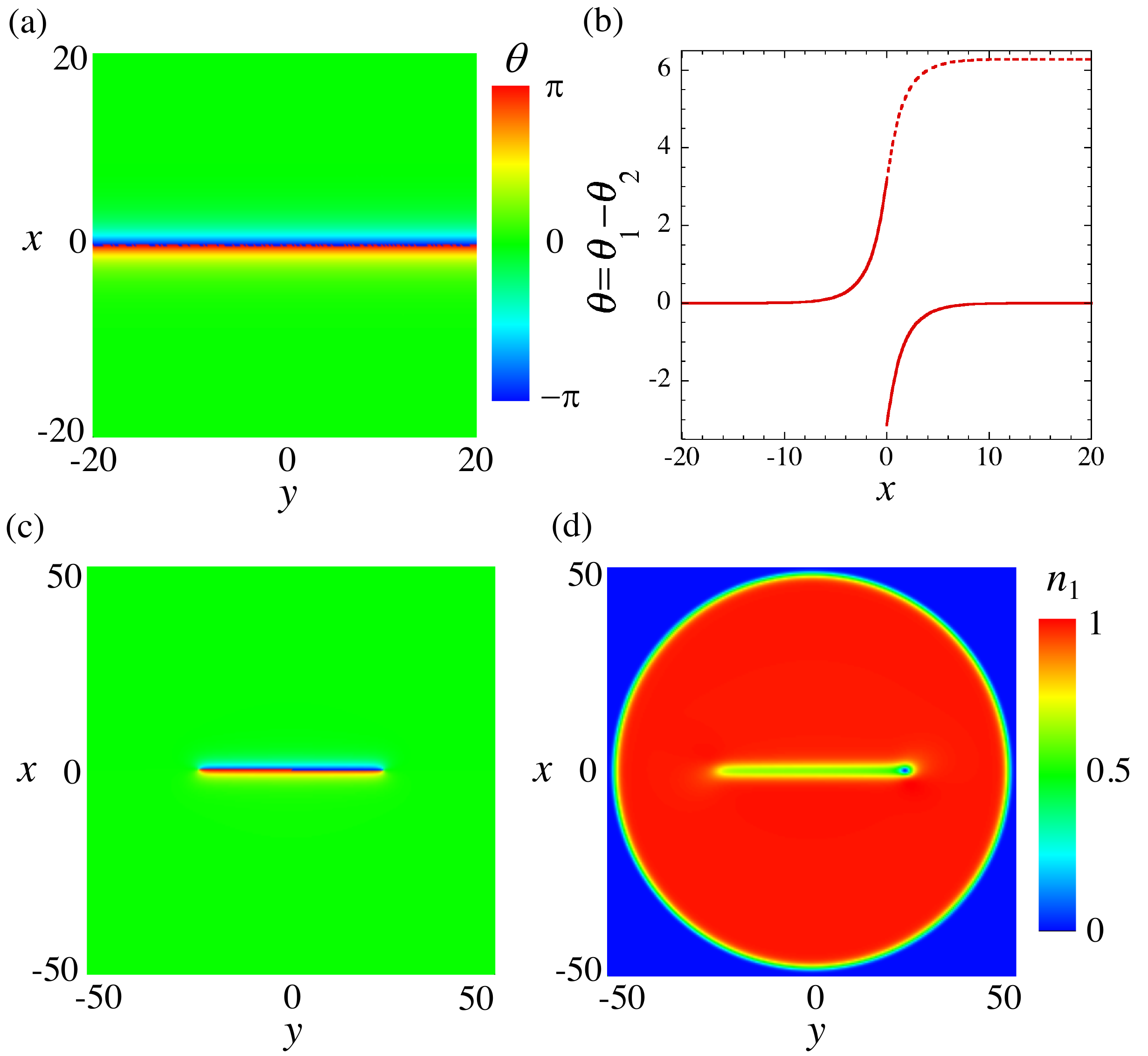}
\caption{The structure of the domain wall of the relative phase in a 2D space. The panels (a) and (b) associate with the situation (i) for $\gamma=0$ and $\omega_\text{R} = 0.15$, 
while (c) and (d) with the situation (ii) for $\gamma=0$, $\omega_\text{R} = 0.1$, and the length $L=45$ in the cylindrical trap of Eq.~\eqref{cylindric}. 
In (c) and (d), a vortex with a unit winding in $\psi_1$- ($\psi_2$-) component is located at $(x,y) = (0, +(-) 22.5)$. 
These states correspond to the initial states of the simulations.  
(a) and (c): The 2D profile of the relative phase $\theta = \theta_1 - \theta_2$. The range of the contourplot is $-\pi \leq \theta \leq \pi$. 
(b): The cross section of $\theta$ in (a) along the $y=0$ line. Here, the solid curves represent $\theta$ within the range $-\pi \leq \theta \leq \pi$. 
The shift of $\theta$ by $2 \pi$ for $x>0$ highlights the structure of the sine-Gordon domain wall as shown by the dashed curve, agreement with the form of Eq.~\eqref{sGsolitonform}. (d): The density profile of $\psi_1$ component, corresponding to (c).  }
\label{molecule}
\end{figure}
Typical structures of the domain wall corresponding to the situations (i) and (ii) are shown in Fig.~\ref{molecule}. 
In the contour plots of $\theta$, we show it within the range $-\pi \leq \theta \leq \pi$ for clarity 
instead of $0 \leq \theta \leq 2\pi$, where we have $2\pi$ phase jump at the center of the wall. 
Then, the wall can be visualized as the localized pattern as shown in Figs.~\ref{molecule}(a) and (c), as done in 
the previous papers \cite{kasamatsu2004vortex,tylutki2016confinement,eto2018confinement}. 
Figure~\ref{molecule} (a) shows the domain wall extended along the $y$-direction, which gives the initial state of the time development in Fig.~\ref{unidynamics}(b). 

This situation (ii) is deeply connected with the experimental observation. 
A straightfoward way to prepare the domain wall of the relative phase in two-component BECs is that 
vortices are prepared in each component and then the internal states are coupled by rf fields to induce 
the Rabi coupling. Then, the vortices are connected with the domain wall of the relative phase. 
Figures~\ref{molecule} (c) and (d) show the typical structure of the vortex molecule. The $\psi_1$-component has a vortex at $(x,y) = (0,22.5)$ with 
the positive unit winding and the $\psi_2$-component has a vortex at $(x,y) = (0,-22.5)$ with the same positive unit winding. 
When the Rabi coupling is applied, the relative phase between two components tends to becomes zero over the entire space.  
However, the phase kinks due to the vortices leave a localized structure in the relative phase between the two vortices, the domain wall being naturally formed. 

\subsection{domain wall in a uniform system}
We first demonstrate the nonlinear dynamics associated with the transverse instability 
through the simulation of the 2D GP equations \eqref{GPlesstwo1} and \eqref{GPlesstwo2} in a uniform system. We first prepare the initial state 
as $\psi_j^\text{ini} (x,y) = \psi_j(x)$ with the solution of Eq.~\eqref{exactsolu}, and calculate the time development using the Crank-Nicholson method. 
The system size is $[-50,50]$ in the $x$-$y$ plane with numerical grids is 1000 $\times$ 1000. We take the Neumann and periodic boundary condition for $x$- and $y$- 
direction, respectively. To initiate the dynamical instability, we add a small random noise $\sim 10^{-6}$ to the initial wave functions. 
We confirm that the domain wall in the stable region in Fig.~\ref{sgphased} is certainly stable in the real time development. 

\begin{figure*}[ht]
\centering
\includegraphics[width=1.0\linewidth]{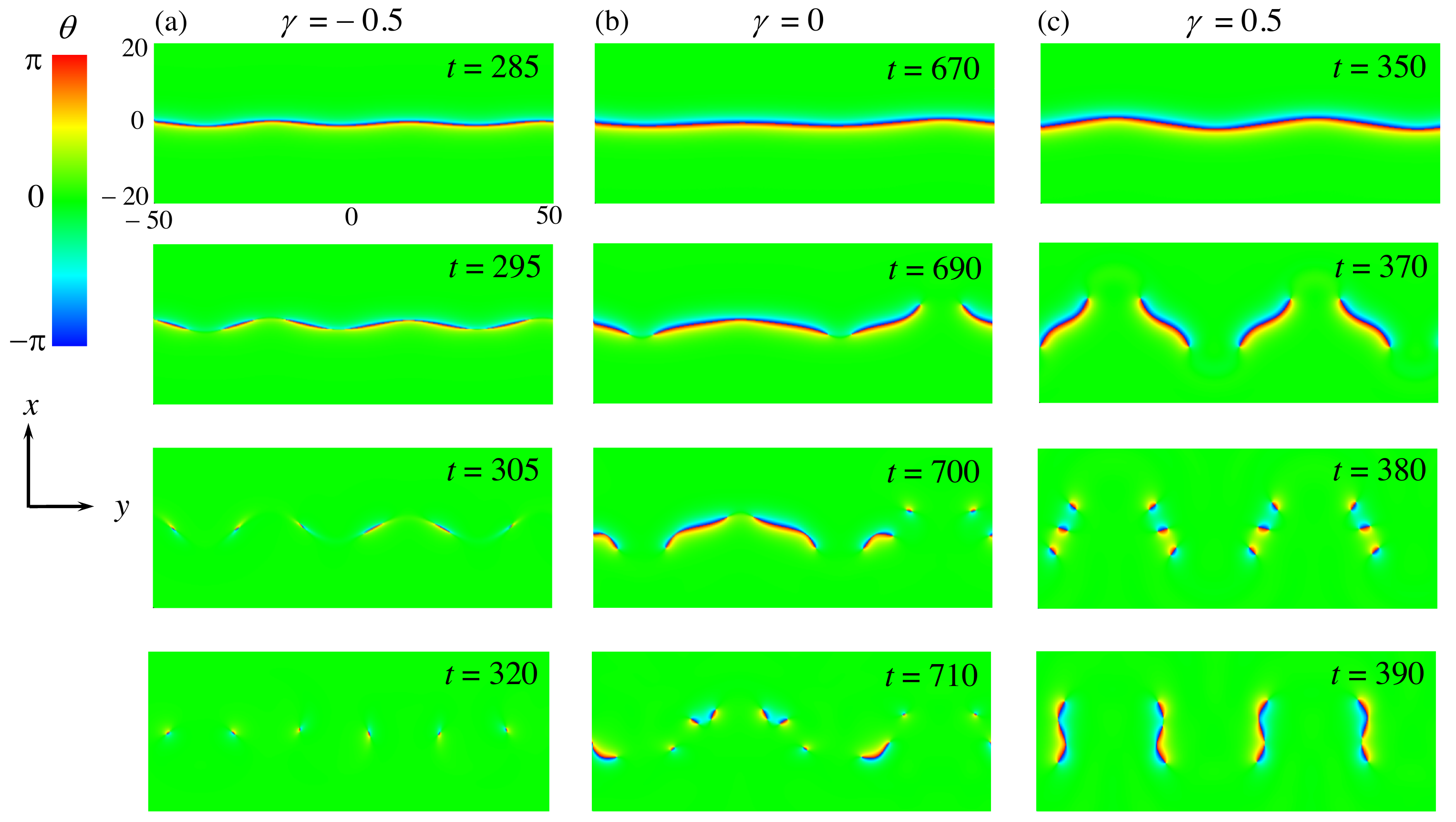}
\caption{The snapshot of the unstable dynamics of the domain wall of the relative phase. 
The panels show the profile of the relative phase defined with the range $-\pi \leq \theta \leq \pi$. 
The parameters are (a) $\gamma = -0.5$ and $\omega_\text{R} = 0.11$, 
(b) $\gamma = 0$ and $\omega_\text{R} = 0.15$, (c) $\gamma = 0.5$ and $\omega_\text{R} = 0.15$. 
We show the profile in the region $-20 \leq x \leq 20$ and $-50 \leq y \leq 50$. }
\label{unidynamics}
\end{figure*}
Figure~\ref{unidynamics} shows the typical snapshots of the disintegration dynamics of the domain wall of the relative phase in the unstable regime for $\gamma=0, \: \pm 0.5$. 
The snapshots show that the small transverse modulation of the wall is amplified after some time and leads to the disintegration of the domain wall. 
According to the BdG analysis, the growing time scale and the wave length of the unstable excitations are estimated 
as $\tau \sim 2\pi/\text{Im}[\omega (k_0)]$ and $\lambda \sim 2\pi/k_0$, respectively; for the parameters in Fig.~\ref{unidynamics}, we have $(\tau,\lambda) = (114.9, \: 32.2)$ for (a), $(\tau,\lambda) = (247.2, \: 64.1)$ for (b), and $(\tau,\lambda) = (149.5, \: 46.5 )$ for (c). 
The simulations results are  in fairly agreement with these estimations; the quantitative deviations may be due to the finite size effect in the simulations. 

After the disintegration, the dynamics of the walls exhibits different behaviors depending on $\gamma$. 
There, the domain wall with the finite size involves the vortices at the edges. 
For $\gamma= -0.5$ in Fig.~\ref{unidynamics}(a), the walls rapidly shrink to zero size due to the combined attractive force 
by the intercomponent coupling and the Rabi coupling between the two components. 
In other words, the vortices at the wall edges attract each other due the attractive vortex-vortex interaction \cite{eto2011interaction} as 
well as the string tension caused by the Rabi coupling. For $\gamma=0$ [Fig.~\ref{unidynamics}(b)], some fragmented walls also tend to shrink 
due to the attraction by the Rabi coupling, but the others keep their separation. 
Such vortex molecules undergo center-of-mass motions, going to outside. 
Eventually, all the walls shrink to zero size at the later stage. 
Contrary to these, the subsequent dynamics for $\gamma=0.5$ is different. Although the initial wall disintegrates into small pieces, 
they keep the disintegration-merge cycle for a while, and after that some walls go to outsides. 
In this case, the repulsive intercomponent interaction $\gamma > 0$ prevents the shrink of the wall. 
In equilibrium, the balance of the repulsive vortex-vortex interaction \cite{eto2011interaction} and the attractive 
force by the Rabi coupling realizes the stable vortex molecule \cite{kasamatsu2004vortex,kasamatsu2005vortices,eto2012vortex,dantas2015bound,shinn2018mesoscopics}.

\subsection{domain wall connecting vortices in a cylindrical trap} \label{cylinsim}
We next consider the dynamics of the domain wall having initially a finite length. 
This situation has been demonstrated in Refs.~\cite{tylutki2016confinement,eto2018confinement}. 
The authors observed that the domain wall rotates due to the attractive tension between the two vortices by the Rabi coupling. 
However, when the separation between vortices or the Rabi coupling becomes large, the wall tends to disintegrate. 
Although, in general, the BdG analysis is not allowed to apply such a non-stationary configuration, we can understand 
qualitatively that their numerical observations are certainly due to the transverse instability as discussed below. 

In simulations, we numerically solve the 2D GP equation with the cylindrical trap 
\begin{equation}
V_\text{ext} = V_0 \left[ 2+\tanh(ar-R) - \tanh(ar+R) \right]  \label{cylindric}
\end{equation}
with the radial coordinate $r$, the potential depth $V_0 = 10$ and the radius $R=50$ of the cylinder. 
The sharpness of the wall boundary is represented by the parameter $a=1$. 
We use the trap of Eq.~\eqref{cylindric} because there is additional contribution to the vortex dynamics from the density inhomogeneity 
when we employ the harmonic trap \cite{tylutki2016confinement,calderaro2017vortex}. 

We simulate the time development with the following procedure. 
First, the initial state without vorticity is prepared for given $\gamma$ and $\omega_\text{R}$ 
through the imaginary time evolution of the GP equation. 
Next, we imprint a vortex in each component by multiplying the phase factor $e^{i\theta_v(\bm{r})}$ with the profile 
$\theta_v (\bm{r}) = \arctan[(y \pm y_0)/x]$ ($-$ for $\psi_1$ and $+$ for $\psi_2$) and make additional imaginary time evolution. 
Then, the vortex separation gradually decreases from $2y_0$, but after some time, the decreasing rate 
reaches a quasi-stationary behavior characterized by the slow linear decrease of the energy, 
which has been also reported in Ref.~\cite{eto2018confinement}. We stop the imaginary time evolution at a certain 
time in this quasi-stationary regime and use this configuration with the vortex separation $L (< 2 y_0)$ as the initial state of the simulation.

\begin{figure}[ht]
\centering
\includegraphics[width=1.0\linewidth]{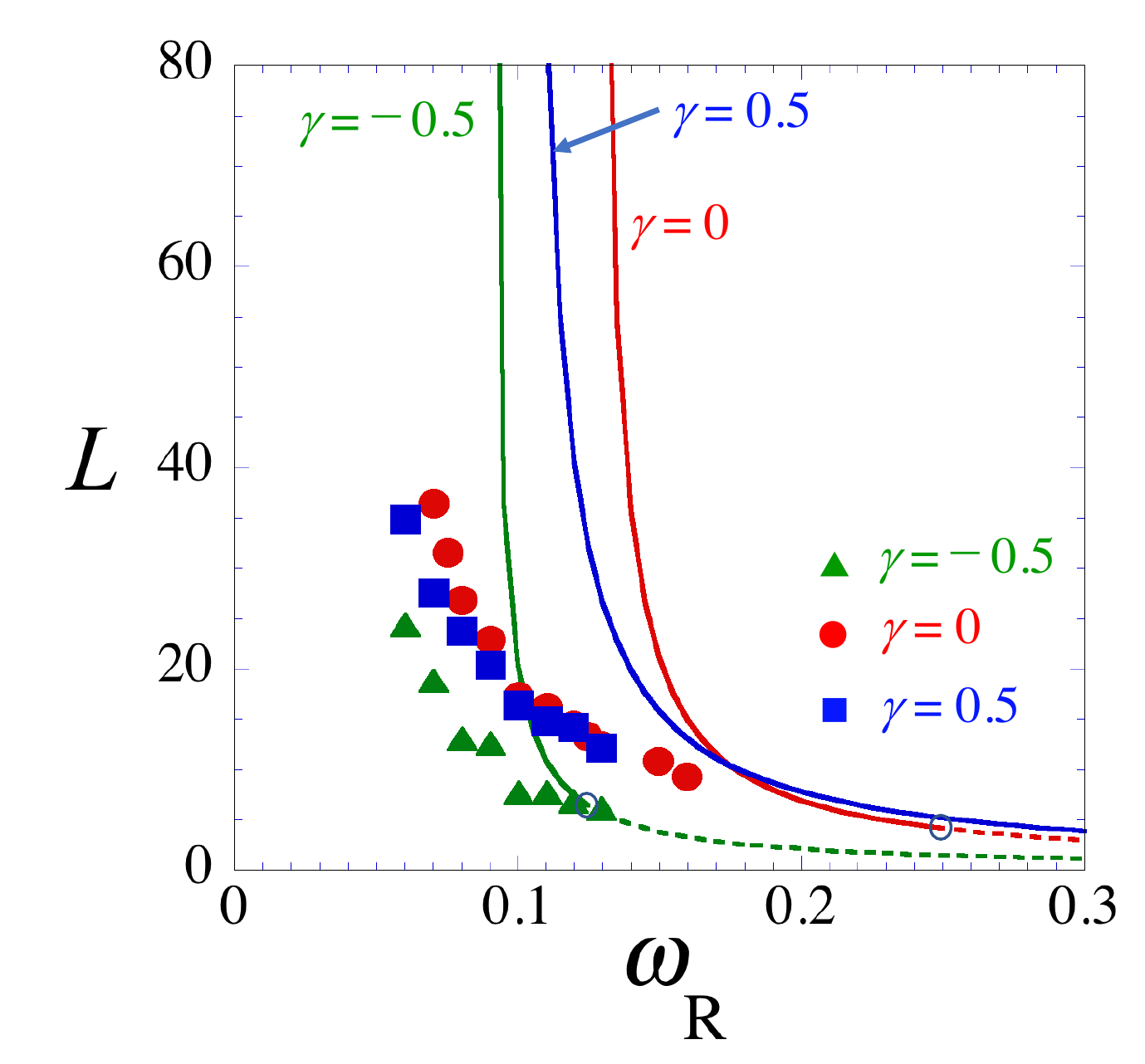}
\caption{The critical length for the disintegration of the domain wall with respect to the Rabi frequency $\omega_\text{R}$. 
The three curves represent the estimation by the BdG analysis for $\gamma = 0, \pm 0.5$, while the plots represent the numerical results. }
\label{Lwphase}
\end{figure}
In this situation, we have to consider the finite size effect for the transverse instability. 
It is necessary for the unstable modes to grow, a half of their wave length should 
be smaller than the length $L$ of the domain wall. 
The condition is given by $\lambda/2 = \pi /k_\text{max} < L$, where $k_\text{max}$ can be written 
as a linear function of $\omega_\text{R}$ [see Fig.~\ref{bdg}(b)]. Using this condition, we get the boundary for the 
disintegration of the finite-size domain wall, as depicted in Fig.~\ref{Lwphase}.
With increasing the Rabi frequency $\omega_\text{R}$, the critical length $L$ of the wall decreases, which is consistent with the observation in Ref.~\cite{tylutki2016confinement,eto2018confinement}.  
The stability diagram is weakly dependent on the intercomponent coupling $\gamma$.

\begin{figure*}[ht]
\centering
\includegraphics[width=1.0\linewidth]{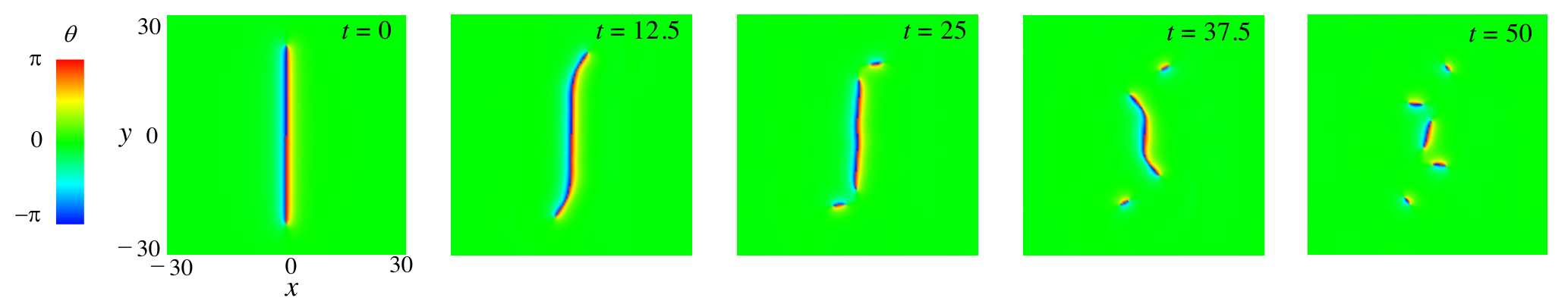}
\caption{The snapshot of the unstable dynamics of the domain wall of the relative phase for $L=45$. 
The panels show the profile of the relative phase defined with the range $-\pi \leq \theta \leq \pi$. 
The parameters are $\gamma = 0$ and $\omega_\text{R} = 0.1$}
\label{dynamics1}
\end{figure*}
Starting from the initial state obtained by the above procedure, we monitor the real time dynamics of the domain wall with the length $L$ for given sets of $\gamma$ and $\omega_\text{R}$.
The typical disintegration process is shown in Fig.~\ref{dynamics1} for $L=45$, $\gamma=0$, and $\omega_\text{R} = 0.1$. 
The wall initially deforms into S-shape and breaks into small pieces at the curved points.  
After that, the remaining wall again deforms into inverted S-shape, repeating the similar breaking process. 
Eventually, the wall breaks into five pieces. 
In this simulation, we take $\gamma=0$ so that there is no repulsive vortex-vortex interaction. 
Thus, the fragmented walls eventually shrink as a result of the tension of the sine-Gordon domain wall. 

When the domain wall disintegrates, the additional vortices are nucleated in each component through the snake instability. 
Thus, monitoring the number of vortices (phase defects) during the time developments provides a clear criterion for the disintegration; 
the number of vortices in each component is kept to be unity when the domain wall is stable, otherwise it is unstable. 
Using this criterion, we calculate the critical length of the wall as a function of $\omega_\text{R}$ for $\gamma=0,\pm 0.5$ and plot in Fig.~\ref{Lwphase}. 
We see that the behavior is qualitatively consistent with the BdG prediction, but quantitatively the disintegration takes place even in the region of the Rabi frequencies smaller than the critical values. 

This quantitative difference may attribute to the rotation of the domain wall with the finite size \cite{tylutki2016confinement,eto2018confinement}.
As mentioned before, the domain wall with the finite size exhibits a rotational motion due to the attractive tension by the Rabi coupling. 
If we consider the rotating frame co-moving with the domain wall, it is seen as a static configuration in this frame. 

To see the stability of moving domain wall in a simple situation, we first consider the energetic stability 
of a moving domain wall with a constant velocity $V$ along the $x$-direction. 
In the co-moving frame of the velocity $V$, the GP equation reads 
\begin{equation}
i  \frac{\partial \psi_j}{\partial t} =  - \frac{\partial^2 \psi_j}{\partial x^2} 
- \tilde{\mu} \psi_j + |\psi_j|^2 \psi_j + \gamma |\psi_{\bar{j}}|^2 \psi_j 
- \omega_\text{R} \psi_{\bar{j}} + i V \frac{\partial \psi_j}{\partial x}. \label{GPvelo1} \\
\end{equation}
Here, the velocity $V$ is scaled by the sound velocity $\xi/\tau$. 
Using Eq.~\eqref{GPvelo1}, we analyze the energetic stability of the domain wall solution 
satisfying the boundary condition similar to Sec.~\ref{exactsolsec} through the imaginary time evolution. 
We show the stability boundary for $V=0.2$ and $V=0.4$ in Fig.~\ref{sgphased}; the stability 
region becomes narrower as $V$ increases. 

Next, we turn to the rotating frame with the frequency $\Omega_\text{prec}$. When we suppose the 
domain wall along the $y$-axis and neglect the influence due to the vortices at the edge of the wall, 
we can approximately employ Eq.~\ref{GPvelo1} with the $y$-dependent velocity $V = - \Omega_\text{prec} y$ 
to discuss the stability problem. 
The precession frequency of the vortex molecule is approximately given by 
$\Omega_\text{prec} \simeq 8 \sqrt{2 \omega_\text{R}}/(\pi L)$ \cite{tylutki2016confinement}.
For $\omega_\text{R} = 0.1$ we find the velocity of the wall at the edge is $V_\text{prec} \sim 0.5$; 
the intermediate region of the wall precesses with the velocity below 0.5. 
Thus, when we take account of the rotational motion of the domain wall, the stability is significantly 
reduced for outer region away from the center of the wall. 
This behavior is clearly seen in Fig.~\ref{dynamics1}, where the disintegration occurs first near the edges.

\section{Conclusion}\label{condle}
We study the transverse instability and disintegration dynamics of the domain wall of the relative phase in 
Rabi-coupled two-component BECs, motivated by the numerical observation in Refs.~\cite{tylutki2016confinement,eto2018confinement}. 
Using the exact solutions, we construct the phase diagram representing the stability of the domain wall. 
In the unstable regime, the domain wall exhibits dynamical instability associated with the transverse displacement, known as the snake instability. The instability causes the disintegration of the domain wall into the small pieces, namely the vortex molecules. 
The growth time and the size of the resulting vortex molecules are in fairly agreement with the prediction by the BdG analysis.

\acknowledgements
The work of K.K. is supported by KAKENHI from the Japan Society for the Promotion of Science (JSPS) Grant-in- Aid for Scientific Research (KAKENHI Grant No. 18K03472). 

\bibliographystyle{apsrev4}
\let\itshape\upshape
\bibliography{reference}

\end{document}